\documentclass[twocolumn,showpacs,aps,floatfix,amsmath,amssymb]{revtex4}
\usepackage{graphicx}% Include figure files
\usepackage{dcolumn}% Align table columns on decimal point
\usepackage{bm}% bold math
\begin{document}
%\preprint{APS/123-QED}%
\title{A Low-Noise High-Density Alkali Metal Scalar Magnetometer}
\author{S.~J.~Smullin}
\author{I.~M.~Savukov}
\author{G.~Vasilakis}
\author{R.~K.~Ghosh}
\author{M.~V.~Romalis}
\affiliation{Physics Department, Princeton University, Princeton, NJ
08544, USA} \pacs{07.55.Ge,32.80 Bx,33.35+r,76.60-k}
\date{\today}
\begin{abstract}
We present an experimental and theoretical study of a scalar atomic
magnetometer using an oscillating field-driven Zeeman resonance in a high-density
optically-pumped potassium vapor. We describe an experimental implementation of an
atomic gradiometer with a noise level below 10~fT~Hz$^{-1/2}$, fractional field sensitivity below
$10^{-9}$~Hz$^{-1/2}$, and an active measurement volume of about 1.5~cm$^3$. We show that the fundamental field sensitivity of a scalar magnetometer is determined by the rate of alkali-metal spin-exchange collisions even though the resonance
linewidth can be made much smaller than the spin-exchange rate by pumping most atoms into a stretched spin
state.
\end{abstract}
\maketitle
\section{Introduction}
High-density hot alkali-metal vapors are used in such vital metrology applications as
atomic clocks \cite{KitchingCl} and magnetometers
\cite{Groeger,AlexMx,BudkerScalar}. In these applications the resolution
of frequency measurements of the hyperfine or Zeeman resonance can
be improved by increasing the density of alkali-metal atoms until
the resonance begins to broaden due to alkali-metal spin-exchange
(SE) collisions. Such broadening can be completely eliminated for
Zeeman resonances near zero magnetic field
\cite{HapperTang,HapperTam,Allred}. The broadening of the hyperfine  and
Zeeman resonances  at a finite magnetic field can be reduced by
optically pumping the atoms into a nearly fully polarized state
\cite{Appelt,Jau04,rfmagn}. These techniques have been used to
demonstrate clock resonance narrowing \cite{Jau04} and have led to
significant improvement in the sensitivity of atomic magnetometers
\cite{KominisRomalisNature} and to their application for detection of magnetic
fields from the brain \cite{Xia} and nuclear quadrupole resonance
signals from explosives \cite{LeeNQR}. However, the effects of SE
collisions on the fundamental sensitivity of magnetometers operating in a finite magnetic field
and on atomic clocks have not been analyzed in detail.
Here we study experimentally and theoretically the effects of SE
collisions in an atomic magnetometer operating in geomagnetic field
range. It was shown in \cite{Appelt,Jau04,rfmagn} that in the
limit of weak excitation the Zeeman and hyperfine resonance
linewidths can be reduced from $\Delta \omega \sim R_{se}$, where
$R_{se}$ is the alkali-metal SE rate, to $\Delta \omega \sim (R_{se}
R_{sd})^{1/2}$, where $R_{sd}$ is the alkali-metal spin-destruction
rate, by pumping most of the atoms into the stretched spin state
with maximum angular momentum. Since for alkali-metal atoms $R_{sd}
\ll R_{se}$ (for example, for K atoms $R_{sd}\sim 10^{-4} R_{se}$),
this technique can reduce the resonance linewidth  by a factor of
$10-100$. However, the frequency measurement sensitivity depends not
only on the linewidth but also on the amplitude of the spin
precession signal, and the optimal sensitivity is obtained for an
excitation amplitude that leads to appreciable rf broadening. In this paper, we
study rf broadening in the presence of non-linear evolution due to
SE collisions and find that the fundamental limit on sensitivity is
determined by $R_{se}$ even when most atoms are pumped into the
stretched spin state and the resonance linewidth is much narrower
than $R_{se}$. We derive a simple relationship for the ultimate
sensitivity of a scalar alkali-metal magnetometer, which also
applies qualitatively to atomic clocks. We find that
the best field sensitivity that could be realized with a scalar alkali-metal magnetometer
is approximately 0.6 fT/Hz$^{1/2}$ for a measurement volume of 1~cm$^3$.

Scalar magnetometers measure the Zeeman resonance frequency
proportional to the absolute value of the magnetic field and can operate in Earth's magnetic field.
They are important in a number of practical applications, such as mineral
exploration \cite{Nabighian}, searches for archeological artifacts
\cite{David} and unexploded ordnance \cite{Nelson}, as well as in
fundamental physics experiments, such as searches for a CP-violating
electric dipole moment \cite{Groeger}. Some of these applications require
magnetometers that can measure small ($\sim$~fT) changes in
geomagnetic-size fields with a fractional sensitivity of $10^{-10} -
10^{-11}$. Existing sensitive scalar magnetometers use large cells
filled only with alkali-metal vapor and rely on a surface coating to
reduce relaxation of atoms on the walls
\cite{AlexMx,BudkerScalar,Groeger}. Here we use  helium buffer gas to reduce diffusion
of alkali atoms to the walls, which also allows independent
measurements of the magnetic field at several locations in the same
cell \cite{KominisRomalisNature}. We present direct measurements of the magnetic
field sensitivity in a gradiometric configuration and demonstrate
noise level below 10 fT ${\rm Hz}^{-1/2}$ in a $10^{-5}$~T static
field (1 part in $10^9$) using an active measurement volume $V \sim
1.5$~cm$^3$. A small active volume and the absence of delicate surface
coatings opens the possibility of miniaturization and batch
fabrication \cite{KitchingMag} of ultra-sensitive magnetometers. The
best previously-reported direct sensitivity measurement for a scalar
magnetometer, using a comparison of two isotopes of Rb occupying the
same volume $V=180~{\rm cm}^3$, had Allan deviation that corresponds to
sensitivity of 60 fT~Hz$^{-1/2}$ and fractional sensitivity of
$5\times 10^{-8}$~Hz$^{-1/2}$ \cite{alexscalarmag}. Theoretical
estimates of scalar magnetometer sensitivity based on photon shot noise level on the order of 1
fT~Hz$^{-1/2}$ have been reported in cells with $V\sim 1000$ cm$^3$
\cite{AlexMx,BudkerScalar}.

We rely on a simple magnetometer arrangement using optical pumping
with circularly-polarized light parallel to the static magnetic
field $B_z$, excitation of spin coherence with an oscillating
transverse magnetic field $2 B_1$, and detection of spin coherence
by optical rotation of a probe beam orthogonal to the static field. RF broadening of
magnetic resonance is usually described by the Bloch equations  with
phenomenological relaxation times $T_1$ and $T_2$ \cite{abragam}.
Since SE collisions generally cause nonlinear spin evolution, such
a description only works for small spin polarization
\cite{BhaskarHapperrfbroad}.  To study the general case of large
polarization and large rf broadening we performed measurements of
resonance lineshapes in K vapor for a large range of SE rates,
optical pumping rates, and rf excitation amplitudes.
We also developed a program for numerical density matrix modeling of the system. To understand the fundamental limits of the magnetometer sensitivity, we derive an analytical result that gives an accurate description of magnetometer behavior in the regime $R_{se} \gg R_{op} \gg R_{sd}$, where $R_{op}$ is the optical
pumping rate, applicable to high density alkali-metal magnetometers with high spin polarization. In the limit of high polarization, we find an implicit equation for the transverse spin relaxation $T_2$ that can be solved to calculate polarization ${\bm P}$ as a function of rf field detuning and other parameters. In this limit, the system is well-described by the solutions to the familiar Bloch equations, with $T_2$ varying as a function of polarization and rf field de-tuning. This modified Bloch equation model reproduces the
non-Lorentzian resonance lineshape from the full density matrix
simulation and the experimental rf
broadening data and allows us to set analytical limits on the magnetometer sensitivity. The same approach can also be easily applied
to other alkali metal atoms with different nuclear spin values and to hyperfine clock transitions.

This paper is organized as follows: Sec.~\ref{sec:measurements} describes the experimental setup and presents measurements of magnetic field sensitivity and other experimental parameters. Sec.~\ref{sec:dynamics} presents a theoretical description of the magnetometer signals. Sec.~\ref{sec:sensitivity} gives expressions for the fundamental sensitivity of the magnetometer and compares this theoretical result to our high-sensitivity magnetometer measurements.
\begin{figure}
\includegraphics[width=0.9\columnwidth]{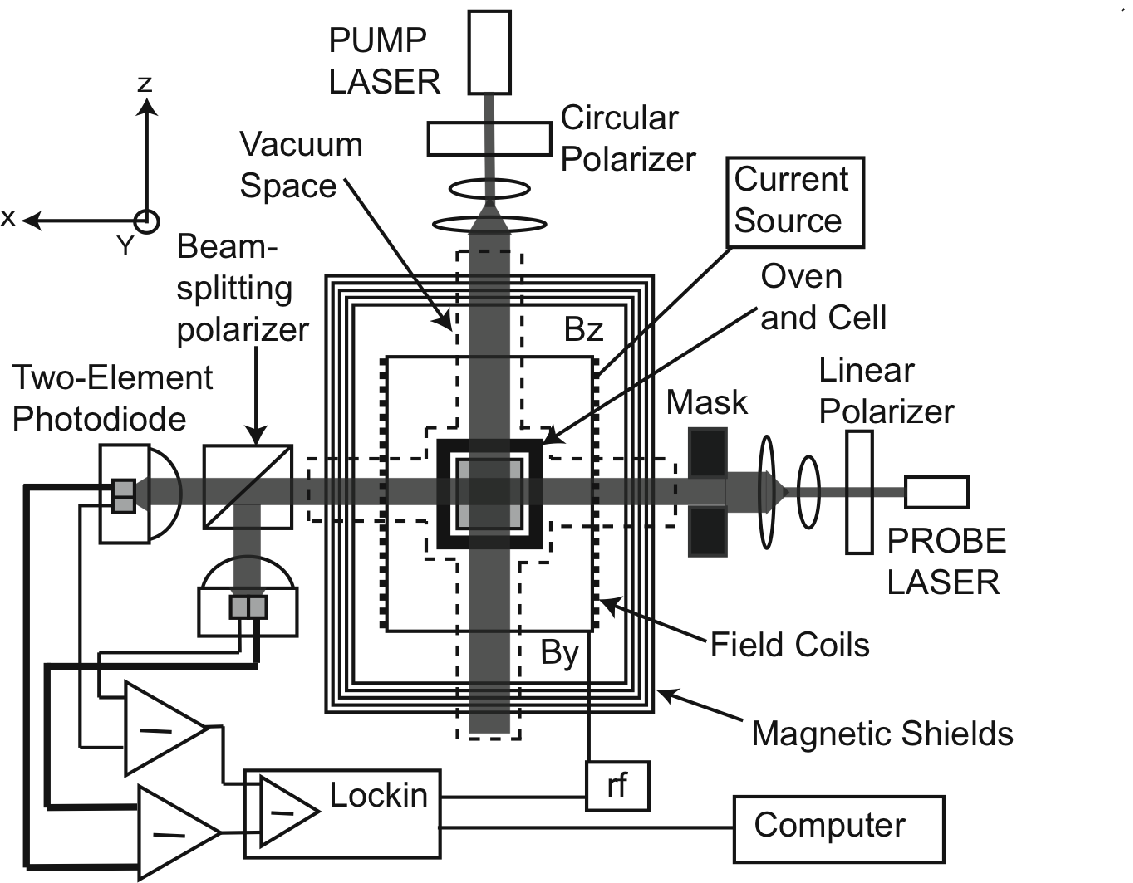}
\caption{\label{fig:expt}Schematic of the experimental apparatus.
The cell ($3\times 3\times 4$ cm$^3$ with the larger dimension
perpendicular to lasers) is heated inside a boron nitride oven and
placed in a glass vacuum enclosure
pumped out to 0.5 Torr. Coils
inside 6-layer magnetic shields allow application of magnetic fields
and gradients. The gradiometer measurement is obtained by imaging
the probe beam onto two-element photodiodes. The signals of the two
balanced polarimeters are subtracted at the lock-in.}
\end{figure}
\section{Experimental measurements}\label{sec:measurements}
\subsection{Measurement apparatus}
The scalar magnetometer, diagrammed in Fig.~\ref{fig:expt}, is built around a Pyrex cell containing
potassium in natural abundance, 2.5 amg of $^4$He to slow atomic
diffusion, and 60 Torr of N$_2$ for quenching. For characterization,
the cell was heated to varying temperatures using a hot air oven.
For the most sensitive magnetometry measurements, the cell was
heated with pairs of ohmic heaters
(wire meander in Kapton sheet) oriented to cancel stray fields and
driven at 27 kHz. A circularly polarized pump beam at the $D1$
resonance polarizes the K atoms along the $z$-direction. The $x$
component of atomic spin polarization is measured using optical
rotation of a linearly-polarized beam as determined by a balanced
polarimeter. Two-segment photodiodes were used on each arm of the
polarimeter to make a gradiometer measurement.
A constant bias field $B_z$ is applied parallel to the pump laser. An oscillating rf field $2B_1$ is applied in the $y$ direction with its frequency tuned to the Zeeman resonance given by
$\omega_0=\gamma B_z=g_s\mu_B B_z/(2I+1)\hbar=2\pi\times (700~{\rm
kHz/G})B_z$ for potassium atoms.  The polarimeter measurement is
read through a lock-in amplifier, tuned to the rf frequency. The
lock-in phase is adjusted to separate the resonance signal into
symmetric (in-phase) absorption and antisymmetric (out-of-phase) dispersion components. Exactly on resonance the dispersive part of the signal crosses zero. The magnitude of the local field is determined by the frequency of this zero-crossing and changes in the dc magnetic field are registered as deviations from zero of the dispersive signal. %By changing the rf frequency and measuring the signal as a function of frequency, we can trace out the Zeeman resonance; typical data are shown in Fig.~\ref{fig:lorentz}.
\subsection{Noise measurements with a high sensitivity atomic magnetometer}\label{lownoise}
Magnetometer noise is read on the dispersive component of the lock-in reading. The conversion of the voltage noise to magnetic field noise depends on the slope as a function of the magnetic field or frequency of the dispersion curve. The tunable parameters of the experiment were adjusted to maximize the dispersion curve slope.
The pump beam (20--40 mW) was imaged on an area of roughly $3\times 1.5$ cm$^2$ across the cell.
A probe beam cross section of  $1.2\times 1.2$ cm$^2$
was defined by a mask with total power of 10 mW and the wavelength detuned by about 100 GHz  from
the $D1$ resonance. After passing through the cell and the polarizing beam splitter the probe beam was
imaged onto two-segment photodiodes. For the most sensitive measurements, the amplitude of the oscillating rf field was about 19 nT.
Magnetic field sensitivity was measured for three values of $B_z$: 1 $\mu$T, 10
$\mu$T, and 26 $\mu$T. The cell was heated to approximately
150$^{\circ}$C, yielding an atomic density
of $n=6.4\times 10^{12}$~cm$^{-3}$.
\begin{figure}
\includegraphics[width=\columnwidth]{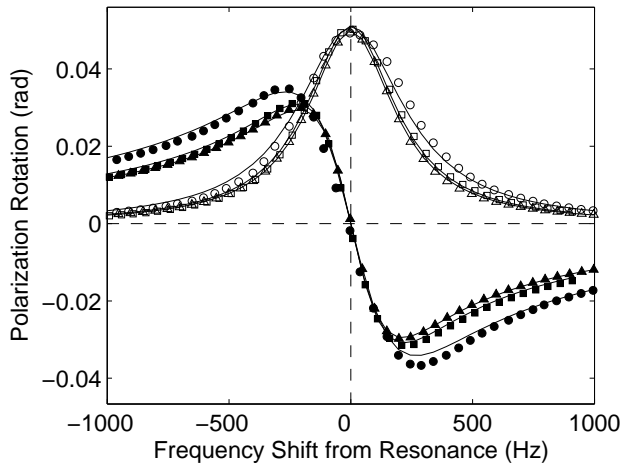}
\caption{\label{fig:lorentz}Absorptive (open symbols) and dispersive
(closed symbols) components of the magnetic resonance polarization
rotation signal at 1 $\mu$T (squares), 10 $\mu$T (triangles), and 26
$\mu$T (circles). Solid lines show Lorentzian fits to the data. These data were recorded at the same time and under the same experimental conditions as the high sensitivity magnetometer measurements.}
\end{figure}
The polarimeter signals were measured with a lock-in amplifier
(Stanford Research Systems SR830 for 1 $\mu$T and 10~$\mu$T measurements,
SR844 for the 26 $\mu$T measurement). The lock-in internal reference
generated the rf field and the time constant was set to
100~$\mu$s. The resonance lineshapes obtained by varying the rf
frequency are shown in Fig.~\ref{fig:lorentz}. The pump power and rf
amplitude are adjusted to optimize the slope of the dispersion
signal for a given probe beam power. At the parameters that optimized
the magnetometer sensitivity, the resonance curves are well-described by Lorentzian lineshapes with similar half-width at half maximum (HWHM) for absorptive and dispersive
components of $\sim$ 220~Hz for 1~$\mu$T and 10~$\mu$T and 265~Hz for 26~$\mu$T. The amplitude and width  of the optical rotation signal was found to be nearly independent of the static magnetic field values over the range of our measurements.
The field $B_z$ was generated using a custom current source, based on
a mercury battery voltage reference and a FET input stage followed
by a conventional op-amp or a transistor output stage
\cite{currentsource}. The fractional current noise was less than
$2\times10^{-8}$ Hz$^{1/2}$ at 10 Hz, about 10 times better than from a
Thorlabs LDC201 ULN current source. Low-frequency ($<10$ Hz) optical
rotation noise was reduced by an order of magnitude by covering the
optics with boxes to reduce air convection that causes beam steering. The oven and laser
beams within the magnetic shields were enclosed in a glass vacuum
chamber to eliminate air currents.

Probe beam position was adjusted to
equalize the photodiode signals for the two polarimeters within $2\%$.
The gradiometer measurements reduced by more than an order of
magnitude the noise from the $B_z$ current source as well as pump
intensity and light shift noise. By applying a calibrated magnetic
field gradient, we found the effective distance between the gradiometer
channels to be $\sim 3.5$~mm, much larger than the K diffusion
length in one relaxation time $(D T_2)^{1/2}\approx0.1$~mm, so the
two measurements are independent.
 % it is really 3.4 for 260 mG, 3. 7 for 10 mG, 3.4 for 100 mG on 6/15 and 6/16.%
%The low power current source
%used for the 10 mG field is more than 15 times better than the
%Thorlabs LDC201 ULN in the 1--10 Hz bandwidth.
%The high power current source, based on a FET input stage
%\cite{currentsource}, is limited only by Johnson noise, sources the
%100 mG field with noise reduced by roughly a factor of 5 with
%respect to the Thorlabs source in the 1--10 Hz bandwidth.
\begin{figure}
\includegraphics[width=\columnwidth]{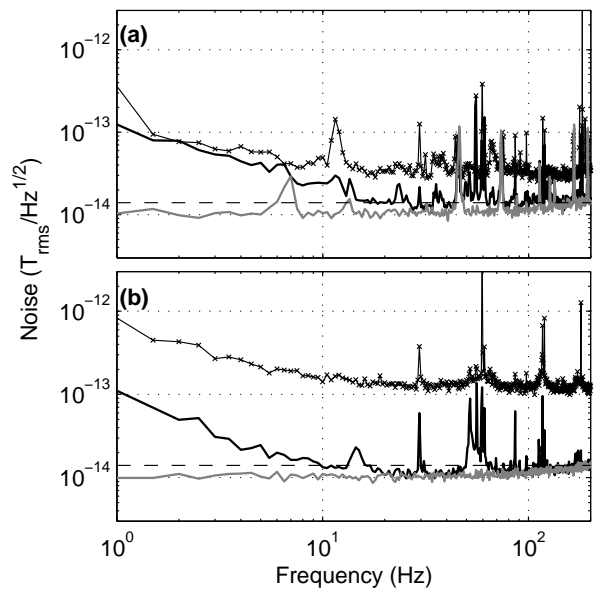}
\caption{\label{fig:results}Noise spectra for 1 $\mu$T (a) and 10 $\mu$T
(b). Shown are single channel spectra (black line with crosses),
two-channel difference (gradiometer) spectra (black solid line), and the measured electronic
and optical noise (gray solid line) obtained by blocking the pump
beam. The dashed black line marks the 14 fT/Hz$^{1/2}$ level.
Magnetic field noise increases at higher frequencies due to
correction for the finite bandwidth of the magnetometer. }
\end{figure}
% real bandwidth:
%for 10 mG, ratio of absorption amplitude: dispersion amplitude= 121/144
%100 mG: 115/133 ; 260 mG: 144/195
%Bandwidths -- I fit the square of the measurement to an absorption of a lorentzian (this should be the same as fitting measurement to magnitude of lorentzian). The half width in this fit (which is the half width of the measurement, divided by sqrt(3) is given below
%10 mG: 190 Hz    100 mG:  201 Hz
% and the half-widths from the Lorentzian fits
%at 260 mG, it doesn't look like such a good fit. Half width of the measurement is 328 Hz. The fit gives a width of 180 Hz. The discrepancy is ok -- this figure is not discussed in paper at this field value, I think.
% 10 Hz fit parameters: 121, 7091,221,-0.37,-144,218
%100 Hz fit parameters: -117,71473,219,3.16,136,215
%260 Hz fit parameters: 144,184000,264,-0.34,-196,267
% this fit parameters are amplitude of absorption, f0, width of absorption, phase offset, amplitude of dispersion, width of dispersion.

The magnetic field data were acquired from the dispersive lock-in
signal for 100 sec with a sampling rate of 2~kHz. The FFT of the
data was converted to a magnetic field noise spectrum using a frequency
calibration of the dispersion slope and corrected for the finite
bandwidth of the magnetometer. The bandwidth was found to be close
to the Lorentzian HWHM for all values of $B_z$. The magnetic noise
spectra at 1 $\mu$T and 10 $\mu$T are shown in Fig.~\ref{fig:results}. At 1
$\mu$T, single channel measurements were limited by lock-in phase
noise, while at 10 $\mu$T they were limited by current source noise.
The noise in the difference of the two channels was limited almost entirely by photon shot
noise at higher frequencies and reached below 14 fT/${\rm
Hz}^{1/2}$, corresponding to less than 10~fT/${\rm Hz}^{1/2}$ for each
individual magnetometer channels. With the pump beam blocked, the
optical rotation noise reached the photon shot noise level. Low
frequency noise was most likely due to remaining effects of
convection. At 26 $\mu$T, the gradiometer had a sensitivity of 29
fT/${\rm Hz}^{1/2}$, limited by lock-in phase noise and imperfect balance between gradiometer channels.
\subsection{Magnetic resonance measurements}
To analyze the magnetometer behavior and predict the theoretical sensitivity of the device, we focus on the shape of the magnetic resonance curves. The basic behavior of the resonance signals can be understood using phenomenological Bloch equations (BE), which predict a Lorentzian resonance lineshape. Though the BE cannot describe the whole physics  in the case of rapid spin-exchange collisions, they do provide a convenient phenomenological framework for qualitative understanding of the resonance lineshape including the effects of rf broadening.  Using the rotating wave approximation, the solution of the BE (see,
for example \cite{abragam}) in a frame rotating about the $z$-axis is:
\begin{eqnarray}
P_i&=&\frac{\Delta \omega \gamma B_1 T_2^2}{1+(\Delta \omega
T_2)^2+(\gamma B_1)^2T_1T_2}P_0 \label{eq:bloch1} \\
P_j&=&\frac{\gamma B_1T_2}{1+(\Delta \omega T_2)^2+(\gamma
B_1)^2T_1T_2}P_0 \label{eq:bloch2} \\
P_z&=&\frac{1+(\Delta \omega T_2)^2}{1+(\Delta \omega T_2)^2+(\gamma
B_1)^2T_1T_2}P_0.  \label{eq:bloch3}
\end{eqnarray}
Here we introduce the in-phase $P_j$ and out-of-phase $P_i$
components of the transverse polarization in the rotating frame and the
longitudinal polarization $P_z$. In the lab frame, we measure $P_x=P_j {\rm cos}(\omega t) + P_i {\rm sin}(\omega t)$, and we tune the lock-in phase to separate the absorptive $P_j$ from the dispersive $P_i$.  $T_1$ and $T_2$ are constant
phenomenological relaxation times, $P_0$ is the equilibrium
polarization, $B_1$ is the amplitude of the excitation field in the
rotating frame, given by $B_y= 2 B_1 {\rm cos}(\omega t)$ in the lab frame. The detuning $\Delta \omega=\omega-\omega_0$ is the difference
between the rf frequency $\omega$ and the resonant frequency
$\omega_0$, which is the Larmor frequency in the applied dc field
$B_z$. The dependencies of $P_i$ and $P_j$ on frequency are
Lorentzian, with the HWHM
\begin{equation} \label{blochrf}
\Gamma=\frac{1}{T_2} \sqrt{1+(\gamma B_1)^2T_2T_1}.
\end{equation}
The increase in the width due to the presence of excitation field $B_1$ is
the basic phenomenon of rf resonance broadening. The slope, at resonance, of the dispersive component of the signal
$dP_i/d\omega(\Delta \omega=0)$ is given by
\begin{equation}
\frac{dP_i}{d\omega}=\frac{\gamma B_1T_2^2}{1+(\gamma
B_1)^2T_1T_2}P_0. \label{slopeBloch}
\end{equation}
The slope has a maximum at an excitation field $B_1=1/(\gamma
\sqrt{T_1 T_2})$:
\begin{equation}
\left.\frac{dP_i}{d\omega}\right|_{\rm max}=P_0 T_2^{3/2}/(2 T_1^{1/2}). \label{max_slope}
\end{equation}

The accuracy of the simple Bloch equation theory depends on the
contribution of spin-exchange relaxation to the linewidth. If the temperature is low, then the
broadening due to optical pumping can exceed SE broadening, and the Bloch
equation theory will be quite accurate. Additionally, if
spin polarization is low, SE broadening will not depend significantly on the polarization and the
excitation field, so the transverse relaxation time $T_2$  will be almost constant; in this case the BE
solution is also valid. However, we are primarily interested in the regime of high spin-exchange rate and high spin polarization, where the magnetometer is most sensitive.

\begin{figure}
%[tbp]
\centerline{\includegraphics[width=0.8\columnwidth]{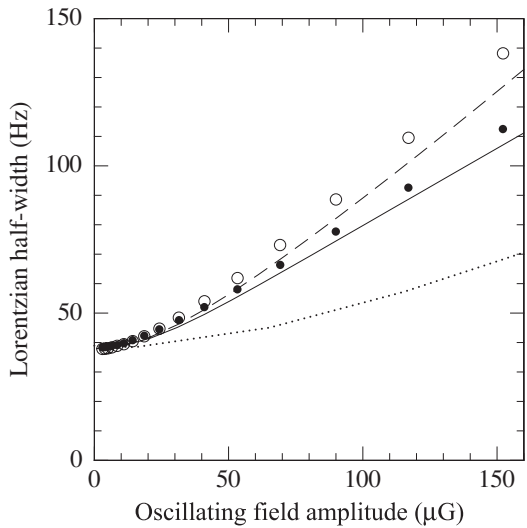}}
\caption{The linewidths of Lorentzian fits to the
experimental data for absorption (solid points) and dispersion (open
points) components of the magnetic resonance in K vapor at
140$^{\circ}$C. The dotted line is the prediction for rf broadening
of the linewidth from Bloch equations with constant $T_1$ and $T_2$, the solid
and dashed lines are results of Lorentzian fits to absorption and
dispersion lineshapes obtained using  modified BE with variable
$T_2$ discussed discussed in the text. Here $R_{se}=5100$ s$^{-1}$ and $R_{sd}=24$ s$^{-1}$ are
fixed from independent measurements, while $R_{op}=840$~s$^{-1}$ is
adjusted to fit the measured linewidth at low rf amplitude.}
\label{linewidthbroad}
\end{figure}

To understand the effects of SE broadening, we compared the lineshape predicted from the BE to the measured resonance lineshape of the magnetometer signal at a frequency of 80 kHz. We recorded the magnetic resonance curves for different values of rf excitation amplitude, pump laser intensity, and cell temperature. We find that the lineshape of the resonance remains reasonably close to a Lorentzian and the in- and out-of-phase lock-in data from resonance measurements were fit to the absorptive and dispersive Lorentzian profiles, allowing for some mis-tuning of the lockin phase.  It can be seen from the BE that relative amplitudes of absorptive and dispersive components  can differ
substantially from those expected from a complex Lorentzian
$1/[i(\omega-\omega_0)+\Gamma]$ in the regime of large rf broadening. Moreover, due to SE effects, the
absorption and dispersion widths for the same experimental
conditions can also differ. Thus, a total of five parameters were used for each resonant curve:
the resonant frequency, and the respective amplitudes and widths of the absorptive and dispersive
signal components. An example of the results for the resonance linewidths as a function of the magnitude of the rf field
is shown in  Fig.~\ref{linewidthbroad}. It can be seen that rf broadening is greater than what is expected from the BE. Moreover, the absorptive and dispersive parts of the resonance have different widths as the rf amplitude in increased. These are signatures of the SE broadening that require modifications of the BE description.

In the regime of small rf broadening we verified that the absolute size of the lock-in signal is in agreement with Bloch equations. The optical rotation signal detected by the lock-in is given by
\begin{equation}
\phi=\frac{lr_ecfn D(\nu) P_i}{2\sqrt{2}}, \label{Rot}
\end{equation}
where $D(\nu)=(\nu-\nu_0)/((\nu-\nu_0)^2+\Gamma_\nu^2)$ is the optical dispersion profile of the $D1$ resonance line with linewidth $\Gamma_\nu$ and oscillator strength $f$, $n$ is the density of atoms, $l$ is the length of the cell in the direction of the probe beam, and $r_e=2.8\times 10^{-13}$ cm is the classical electron radius. Here we take into account the fact that lock-in output measures the r.m.s. of an oscillating signal. The length $l$ is determined by the dimensions of polarized vapor illuminated with the pump beam. Near the edges of the cell the pump beam is distorted, reducing $l$ below the inner dimensions of the cell. We varied the width of the pump beam to find that the largest pump width for which the signal still increases is about 2 cm. For this value of $l$ the absolute signal size was in agreement with Bloch equations to within 15\%. So the volume of the polarized atomic vapor participating in the measurement is about $2\times 1.2\times 0.6$ cm$^3 \sim 1.5$ cm$^3$.

\subsection{Relaxation rates}
A number of independent auxiliary measurements were preformed to find the relaxation rates of the alkali-metal spins to be used for detailed modeling of SE effects.  Spin exchange and spin destruction rates can be determined by measuring the width of the Zeeman resonance in a very low field using low pump and probe laser intensity \cite{Allred}.  For these measurements, the magnetic field was perpendicular to the plane of the lasers and the pump laser intensity was modulated near the Zeeman resonance. The signal  as a function of modulation frequency was fit to a sum of two Lorentzians taking into account the counter-rotating component of pump rate modulation ~\cite{MagnPRA}. At low magnetic field, when the Zeeman frequency is much smaller than the SE rate, SE broadening depends quadratically on the magnetic field. The spin-destruction rate is obtained from extrapolation of the width to the zero-field limit.  From these fits of the resonant frequency and linewidth we determined the spin-exchange rate $R_{se}$ and the spin-destruction rate  (SD) $R_{sd}$, which are listed in Table~\ref{tbllow} for the same cell at several temperatures. The error bars are estimated from fits to different sets of the data. In addition, we determined the density of K atoms by scanning the DFB probe laser across the the optical absorption profile of the D1 resonance.  The alkali densities calculated from the integral of the absorption cross-section using known oscillator strength ($f=0.34$) and  cell length ($l=3$ cm) are also shown in Table ~\ref{tbllow}. The density is approximately a factor of 2 lower than the density of saturated K vapor at the corresponding temperature, as we find is common in Pyrex cells, probably due to slow reaction with glass walls. The alkali-metal SE rate can be calculated from the measured density using known K-K spin-exchange cross-section $\sigma_{SE} =1.78\times 10^{-14}$~cm$^{2}$~\cite{erratumse} and is in good agreement with direct measurements. The spin destruction rate $R_{sd}$ can also be calculated using previously measured spin-destruction cross-sections for K-K, K-He and K-N$_2$ collisions~\cite{ChenWalker} and gas composition in the cell (2.5 atm of $^4$He and 60 torr of N$_2$). We also include relaxation due to diffusion to cell walls. Errors on the rates calculated from the densities are estimated from uncertainty in the cross-section and in the gas pressures in the cell. Our direct measurements of the spin-destruction rate are reasonably consistent with these calculations.
\begin{table}
 \caption{Comparison of the measured SD  rates $R_{sd}^{m}$ and SE rates $R_{se}^{m}$
 from fits of the resonance linewidth at low field  with
 corresponding rates $R_{sd}^{cal}$ and $R_{se}^{cal}$ calculated from collision cross sections and the
   density of K metal determined by optical absorption.
 }
  \label{tbllow}
\begin{tabular*}{0.45\textwidth}{@{\extracolsep{\fill}} c c c c c c}
\hline \hline Temp.& $R_{sd}^{m}$&$R_{sd}^{cal}$ & $R_{se}^{m}$
& $R_{se}^{cal}$& K density  \\
$^\circ$C &s$^{-1}$ &s$^{-1}$& ms$^{-1}$&  ms$^{-1}$& 10$^{12}$ cm$^{-3}$  \\
\hline
 130 & 28$\pm$3 & 21$\pm$2 & 2.3$\pm$0.2    & 2.5$\pm$0.1  & 2.2\\
 140 & 22$\pm$2 & 22$\pm$2 & 4.2$\pm$0.2    & 4.5$\pm$0.2  & 3.8\\
 150 & 43$\pm$6 & 23$\pm$2 & 10$\pm$0.5   & 8.1$\pm$0.2 & 6.7 \\
 160 & 31$\pm$3 & 25$\pm$2 & 14$\pm$1.0   & 13.9$\pm$0.2 & 11.4 \\
\hline
\end{tabular*}
\end{table}
% noise comparisons%
% if i compare the sqrt(sum(power)) from 6/28 data
% at 100 mA current,
% my source is factor of 4 better in 1-10 Hz
% and factor of 2 better in 1-100 Hz
% if I subtract (as power) the background noise,
% then I see a factor of 6 improvement with my current source in 1-10 Hz bw.
% if I look at 6/16 measurement "inoise", i see a factor of 9 improvement
% in the 1-10 Hz bw; here I use monitor out; this is more appropriate
% b/c don't have all this resistor noise. It is ok to compare this to
% thorlabs source with resistor since resistor noise small compared
% to current source noise in this case.
% If I use Mike's current source comparisons from 4/4, which are also not
% impeded by resistor noise, at ~12 mA, then I find that compared to thorlabs
% source, his source is ~ factor of almost 18 improvement in the 1--10 Hz bandwidth
% and a factor of almost 10 improvement 1-100 Hz. Note these measurements done with
% 100 Hz bandwidth while my measurements are done in 200 Hz bw.
\section{Model of magnetometer dynamics}\label{sec:dynamics}

We first model the dynamics of the system using numerical evolution of the density matrix to  accurately describe the effects of SE relaxation.  To provide more qualitative insight and estimate the fundamental limits of sensitivity we also develop a semi-analytical description, a modification of the BE, that provides a good approximation to the numerical solutions in the regime of high spin-exchange rate.

\subsection{Density matrix equations}
The spin evolution can be accurately described  by the solution of the
Liouville equation for the density matrix. The time-evolution of the density
matrix $\rho(t)$ includes hyperfine interaction, static and rf field interactions, optical pumping, spin relaxation processes and non-linear evolution due to alkali-metal spin-exchange collisions. In the presence of high density buffer gas when the ground and excited state hyperfine structure of the alkali-metal atoms is not resolved optically, the density matrix evolution is given by the following terms ~\cite{HapperTheory}:
\begin{eqnarray}
\frac{d\rho }{dt} &=&\frac{A_{hf}}{i\hbar }[\bm{I}\cdot \bm{S},\rho ]
+\frac{\mu _{B}g_{S}}{i\hbar}[\bm{B}\cdot \bm{S},\rho ]+\frac{\varphi -\rho }{T_{sd}}\\
&+&\frac{\varphi \left( 1+4\left\langle \bm{S}\right\rangle \cdot \bm{S}%
\right) -\rho }{T_{se}} +R_{op}[\varphi (1+2\bm{s}\cdot \bm{S}%
)-\rho ]. \nonumber  \label{densmatr}
\end{eqnarray}
Here, $A_{hf}$ is the hyperfine coupling, ${\bm I}$ is the nuclear spin and ${\bm S}$ is the electron spin operator. The Bohr magneton is $\mu_B$ and $g_s$ is the electron $g$-factor, $\bm{B}$ is the external magnetic field including static and oscillating components, $\varphi$ is the purely nuclear part of the density matrix ~\cite{HapperTheory}, and ${\bf s}$ is the spin polarization of the pump beam. We evaluate the density matrix in the $|F,m\rangle$ basis and focus on the regime of relatively low static magnetic field, where the non-linear Zeeman splitting given by the Breit-Rabi equation is small. To simplify numerical solution of the non-linear differential equations we neglect hyperfine coherences and make the rotating wave approximation for Zeeman spin precession,
\begin{equation}
\langle F,m|\rho(t)|F',m'\rangle= \delta_{F,F'} \langle F,m|\rho'(t)|F,m'\rangle e^{i \omega (m'-m) t}\label{densharm} .
\end{equation}
Here $\omega$ is the frequency of rf excitation field tuned near the Zeeman resonance and $\langle F,m|\rho'(t)|F,m'\rangle$ is the density matrix element in the rotating frame, evolving on a time scale on the order of spin relaxation rates that are  much slower than the Zeeman spin precession frequency. With this approximation  it is necessary to consider only 21 elements of the density matrix for $I=3/2$ using the symmetry of the off-diagonal components.
In the rotating frame without loss of generality we parameterize the density matrix as
$\rho'(t)=\rho'_{ST}(\beta,\theta,\phi)+\rho'_1 $. Here $\rho'_{ST}(\beta,\theta,\phi)$ is a spin-temperature distribution  $ \langle F,m|\rho'_{ST}(\beta)|F,m'\rangle \propto e^{\beta m}$ that is rotated by an angle $\theta$ from the $z$ axis into the $j$ direction of the rotating frame and an angle $\phi$ around the $z$ axis. $\rho'_1$ is a density matrix describing deviations from spin-temperature distribution, which are small because the spin-exchange rate is much larger than all other rates. For a given value of the spin temperature $\beta$ and angles $\theta$ and $\phi$ the expectation value of $\langle {\bf S}\rangle={\rm Tr}[{\bf S} \rho'_{ST}(\beta,\theta,\phi)]$  is used in the spin-exchange term of the density matrix evolution equations, reducing them to a set of linear first order differential equations for the perturbation matrix $\rho'_1$. The steady-state solution for $\rho'_1$  is obtained symbolically in Mathematica. To obtain a self-consistent solution, $\beta$, $\theta$ and $\phi$ are adjusted until the steady-state solution for $\rho'_1$ satisfies ${\rm Tr}[{\bf S} \rho'_1]=0$. The self-consistency iteration is performed numerically  for various values of the optical pumping rate and the rf excitation strength and detuning.

\subsection{Modified BE}
Though the numerical solutions to the density matrix equations give an accurate treatment of the spin dynamics, it is convenient to develop an analytical model that can describe the asymptotic behavior of the system in the regime of high spin-exchange rate. Here we focus on the regime of light-narrowing \cite{Appelt,rfmagn}, with $R_{se}\gg R_{op} \gg R_{sd}$, which also implies that $P$ is close to unity. For weak rf excitation an analytic expression for $T_2$ under these conditions has been obtained in \cite{HapperTheory,Appelt,rfmagn},
\begin{equation}\label{widthextra}
\frac{1}{T_2}=\frac{R_{op}}{4}+\frac{R_{se}} {5} (1-P_z).
\end{equation}
The coefficients in this expression depend on the nuclear spin $I$ and on the size of the nonlinear Zeeman splitting relative to the spin-exchange rate \cite{rfmagn}. Eq.~(\ref{widthextra}) describes the case of $I=3/2$ and large spin-exchange rate relative to the non-linear Zeeman splitting, so all Zeeman resonances overlap. It is clear from this equation that spin-exchange relaxation can be suppressed by maintaining $P_z$ close to unity.

To extend this solution to arbitrary rf excitation we observe that the relaxation due to spin-exchange and optical pumping is independent of the direction of spin polarization. Therefore, we can apply Eq. (\ref{widthextra}) in a rotating frame with $z'$ axis tilted by an angle $\theta$ from the lab $z$ axis and rotating together with $\mathbf{P}$ in the presence of a large rf excitation field. In doing so we introduce an error due to inaccurate treatment of transverse spin components in the $F=1$ state. Spin precession in $F=1$ state occurs in the direction opposite to the precession in $F=2$ state and hence will not be stationary in the rotating frame. However, this error is small in the light narrowing regime because of two small factors: a) the population in $F=1$ state is small since $P$ is close to unity and most atoms are pumped into the stretched state with $F=2$ and  b) $\theta \ll 1$ for rf fields that provide optimal sensitivity to maintain $P$ close to unity and hence the transverse components of spin are small.

\begin{figure}
%[tbp]
\centerline{\includegraphics[width=0.8\columnwidth]{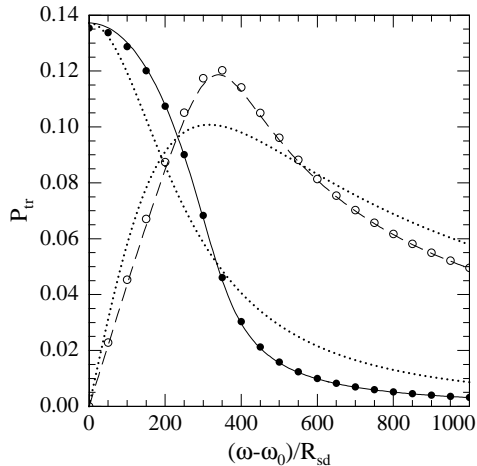}}
\caption{ Comparison of transverse polarization components ($P_i$, $P_j$) using
full numerical density matrix evolution (solid points -- absorption,
open points -- dispersion) and modified BE (solid line --
absorption, dashed line -- dispersion) for $R_{se}/R_{sd}=10^4$,
$R_{op}/R_{sd} = 200$, $\gamma B_1/R_{sd} = 100$. Lorentzian
lineshapes are shown with dotted lines for comparison.}
\label{BlochAnComp}
\end{figure}

Using this approximation we then solve BE (Eq. (\ref{eq:bloch1}-\ref{eq:bloch3})) in combination with an equation for $T_2$ as a function of polarization
\begin{equation}\label{T2corrected}
\frac{1}{T_2}=\frac{R_{op}}{4}+\frac{R_{se}} {5} [1-(P_i^2+P_j^2+P_z^2)^{1/2}].
\end{equation}
The longitudinal spin-relaxation time is not affected by spin exchange and is given by $T_1=4/(R_{op}+R_{sd})$ in the limit of high spin polarization \cite{rfmagn,Appelt}. The equilibrium spin polarization in the absence of rf excitation is equal to $P_0=R_{op}/(R_{op}+R_{sd})$.
The resulting algebraic equations can be easily solved for arbitrary parameters. However, the solution is only expected to be accurate when $P$ remains close to unity. In Fig.~\ref{BlochAnComp} we compare the resonance lineshapes obtained with a full numerical density matrix and the analytical calculation using modified BE. It can be seen that for this case which is well into the asymptotic regime $R_{se}\gg R_{op} \gg R_{sd}$ the analytical results agree very well with exact calculations. The lineshapes are significantly different from a simple Lorentzian.

\subsection{Comparison of experimental measurements with theory}
\begin{figure}
%[tbp]
\centerline{\includegraphics[width=0.8\columnwidth]{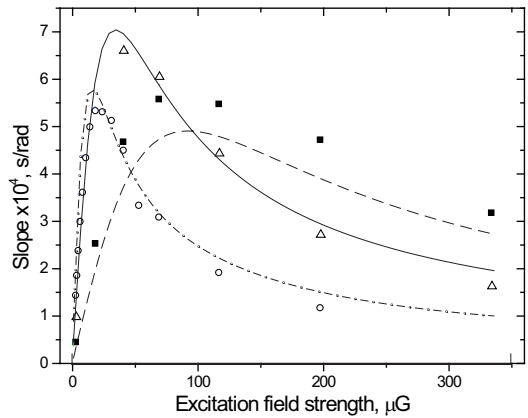}}
\caption{ Slope of the dispersive part of the resonance curve, given as polarization per angular frequency $dP_i/d\omega$. Experiment: Open circles -- $R_{op}=220$ s$^{-1}$,
open triangles -- $R_{op}=625$ s$^{-1}$, solid squares -- $R_{op}=1450$
s$^{-1}$. Theory: dash-dotted line --$R_{op}=220$ s$^{-1}$, solid line --
$R_{op}=625$ s$^{-1}$, dashed line -- $R_{op}=1450$ s$^{-1}$. Temperature
$T=140$ $^\circ$C, $R_{sd}=24$ s$^{-1}$, $R_{se}=5100$ $^{-1}$.
} \label{slope}
\end{figure}

\begin{figure}
%[tbp]
\centerline{\includegraphics[width=0.8\columnwidth]{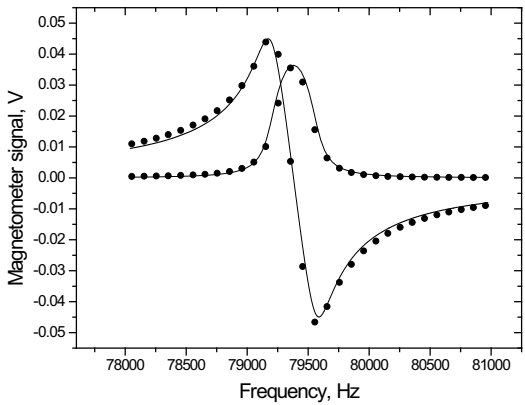}}
\caption{ Comparison of experimental and theoretical resonance lineshapes.
$T=140$ $^\circ$C, $R_{sd}=24$ s$^{-1}$, $R_{se}=5100$
s$^{-1}$, $R_{op}=625$ s$^{-1}$, $B_1=20$ nT. %200 uG%
The model also includes a correction for the polarization-dependent absorption of the pump beam.
%$P_z$ at
%resonance for this condition is 0.263.
} \label{absdispan}
\end{figure}

The results of the simple BE, numerical calculations with the full density matrix equations, and analytical results from the modified BE were compared to a large set of measurements in various parts of the parameter space.  One such comparison is shown in Fig.~\ref{linewidthbroad}. The  experimental data compare well to the BE when a variable $T_2$ (from Eq.~\ref{T2corrected}) is used. Note that not only is the measured width greater than that predicted by the simple BE (with constant $T_2$) but also, as correctly predicted from the analytical theory, the half-width of the absorption curve  differs from the half-width of the dispersion curve. At higher excitation amplitudes, even the modified Bloch analysis begins to deviate from the measured half widths because the polarization begins to drop. The absorbtivity of the vapor also changes as a function of the rf excitation and the pumping rate at the location of the probe beam is not a constant. This can be taken into account by considering the propagation of the pumping light through the polarized vapor. Though the width of the resonance is a good metric for comparing experiment to theory, it is the slope of the dispersive component at resonance that is most important for the magnetometer sensitivity. In Fig.~\ref{slope} is shown a comparison of the measured slopes (from the same data as Fig.~\ref{linewidthbroad}) to those predicted by the analytical theory. The agreement is generally satisfactory. In Fig.~\ref{absdispan} we show one example of a fit of the measured resonance profile to that predicted from the modified BE.   As these data show, in the parameter-space of interest, the modified Bloch analysis provides a good description of the slope and the width of the resonance, as a function of $R_{op}$, $B_1$, $R_{sd}$, and $R_{se}$. Thus we can use these equations to determine the best-achievable sensitivity of the scalar magnetometer.
\section{Magnetometer sensitivity}\label{sec:sensitivity}
For a given slope of the dispersion curve, unavoidable noise sources in the system determine the fundamental sensitivity of the scalar magnetometer. The calculation of the sensitivity follows closely that for an rf atomic magnetometer,
derived in \cite{rfmagn}. For a given polarization noise $\delta P_i$, the resulting field noise $\delta B$ is
\begin{equation}\label{dBP}
\delta B=\frac{\delta P_i}{\gamma |dP_i/d\omega|}.
\end{equation}
There are many sources of technical noise which contribute either
directly to the scalar magnetometer noise as in the case of
low-frequency magnetic field noise from the current source, or indirectly
as in the cases of voltage noise of an amplifier, magnetic field noise at high frequency, vibrations of
the optical detection system, and pump laser noise. Technical noise can
be removed in principle, so it is important to understand the
fundamental limits that determine the best
achievable sensitivity.
\subsection{Photon shot noise}
In a balanced polarimeter the polarization rotation noise per unit bandwidth due to quantum fluctuations of the number of photons received by photodetectors with quantum efficiency $\eta$ is given by
\begin{equation}
\delta\phi=1/\sqrt{2\Phi_{pr}\eta},
\end{equation}
where $\Phi_{pr}$ is the number of photons per second in the probe beam. The noise has a flat frequency spectrum and $\delta\phi$ is measured in units of rad/Hz$^{1/2}$. The same level of noise per unit bandwidth will be measured in each phase of a lock-in amplifier calibrated to measure the r.m.s. of an oscillating signal. The optical rotation measured by the lock-in amplifier is given by Eq.~(\ref{Rot}).

It is convenient to express the photon flux $\Phi_{pr}$ in terms of the pumping rate of the probe beam $R_{pr}$:
\begin{equation}
R_{pr}=\frac{r_e cf\Phi_{pr}(\Gamma_\nu/A)}{(\nu-\nu_0)^2+\Gamma_\nu^2}\label{rpr},
\end{equation}
where $A$ is the cross-sectional area of the probe beam. If the probe laser is detuned far from resonance, $|\nu-\nu_0| \gg \Gamma_\nu$, then one can  express photon-atom interactions in terms of the number of absorption lengths on resonance $N_{ab}= r_e c f n l /\Gamma_\nu$. Using  Eq.~(\ref{dBP}) and magnetometer volume $V=lA$ we find the magnetic field noise due to photon shot noise is
given by
\begin{equation}
\delta B_{PS}=\frac{2}{ \gamma |dP_i/d\omega | \sqrt{N_{ab} R_{pr}nV\eta} } .
\end{equation}

\subsection{Light-shift noise}
The ac Stark shift (light shift) is induced by the probe beam, which is tuned off-resonance
from the atomic transition, when it has a non-zero circular polarization. If the probe laser detuning is much
larger than the hyperfine splitting, the action of the light on atomic spins is equivalent to the
action of a magnetic field parallel to the light propagation direction. This light-shift field
is given by (see Eq.~(9) of Ref.~\cite{rfmagn}),
\begin{equation}
B_x^{LS}=\frac{r_ecf\Phi_{pr}s_xD(\nu)}{(2I+1)\gamma A},
\end{equation}
where  $s_x$ is the degree of circular polarization of the probe beam. Light-shift noise can occur
as a result of fluctuations of intensity, wavelength, or $s_x$. If the probe beam is perfectly linearly polarized,
 fluctuations of the circular polarization are due to quantum fluctuations resulting in an imbalance between
the number of left and right circularly polarized photons in the probe beam. The spectral density of the
probe beam spin polarization noise is given by $\delta s_x=\sqrt{2/\Phi_{pr}}$.
Substituting this value of $s_x$ and excluding $\Phi_{pr}$ by using the pumping
rate of the probe beam $R_{pr}$  in the limit $(\nu-\nu_0)\gg\Gamma_{nu}$ we get:
\begin{equation}
\delta  B_x^{LS}=\frac{ \sqrt{2 R_{pr}N_{ab}} } {4 \gamma \sqrt{nV} }.
\end{equation}
This effective field noise ($B_x^{LS}\ll B_1$) causes polarization noise by rotating the $P_z$ component into the direction of the primary signal $P_i$.  The  amount of polarization noise in $P_i$ induced by the light-shift field is proportional to the spin coherence time $T_2$. Using simulations of BE with noise terms one can verify that
\begin{equation}
P_i^{LS}=\gamma B_x^{LS} T_2 P_z /\sqrt{2}
\end{equation}
where a factor $1/\sqrt{2}$ appears because only the component of the light-shift
field that is co-rotating with the spins contributes to the noise. We get the following contribution of the light shift to the
noise of the magnetometer
\begin{equation}
\delta B_{LS}=\frac{P_z\sqrt{R_{pr}N_{ab} T_2^2}}{4 \gamma \sqrt{nV} |dP_i/d\omega |} .
\end{equation}
In most cases of interest here one can assume that $P_z\simeq1$.

\subsection{Spin projection noise} The spin-projection noise occurs as a result of quantum fluctuations in the components of atomic angular momentum. We consider the case when the polarization is close to unity and most atoms are in $F=2$ state. Using the fundamental uncertainty relationship $\delta F_x \delta F_y\geq\hbar F_z/2$ with $\langle F_z \rangle\simeq2$ one can show \cite{rfmagn} that the polarization noise per unit bandwidth is given by
\begin{equation}
\delta P_i=\sqrt{T_2/N} \label{dfx},
\end{equation}
where $N$ is the total number of atoms. The spin projection noise depends only weakly on absolute spin polarization; for K atoms with $I=3/2$, it increases by $\sqrt{3/2}$ for unpolarized atoms. The resulting magnetic field noise in the scalar magnetometer is given by
\begin{equation}
\delta B_{SP}=\frac{\sqrt{T_2/N}}{\gamma |dP_i/d\omega
|} .
\end{equation}
\subsection{Optimization of fundamental sensitivity}
Combining all the noise contributions we obtain the following equation for the magnetometer sensitivity
\begin{equation}
\delta B=\frac{\left[ \frac{dP_i}{d\omega }\right] ^{-1}}{\gamma \sqrt{nV}}\sqrt{T_2+\frac{T_2^2 R_{pr}N_{ab}}{16}+%
\frac{4}{R_{pr}N_{ab}\eta }} \label{unoptnoise},
\end{equation}
The first term describes spin projection
noise, the second, the light shift of the probe beam, and the third,
photon shot noise.

To find the fundamental limit of the sensitivity we assume that $N_{ab}$ can be adjusted separately, for example by increasing the length of the sensing region in the probe direction while keeping the volume constant, or changing the buffer gas pressure. We find that the optimal optical length is equal to  $N_{ab}=8/(\sqrt{\eta}T_2 R_{pr})$. It is always beneficial to reduce $R_{pr}$ and increase $N_{ab}$, which will result in longer $T_1$ and $T_2$ until $R_{pr}\ll R_{sd}$. Under optimal probing conditions the fundamental magnetometer sensitivity reduces to
\begin{equation}
\delta B=\frac{\left[ \frac{dP_i}{d\omega }\right] ^{-1}}{\gamma
\sqrt{nV}}\sqrt{T_2(1+\eta^{-1/2})}.
\label{optOD}
\end{equation}

The best sensitivity is obtained by maximizing $dP_i/d\omega/\sqrt{T_2}$. For a given
$R_{se}$ and $R_{sd}$ we vary $R_{op}$ and $B_1$ and calculate $dP_i/d\omega$ and $T_2$ using modified BE
with variable $T_2$ given by Eq.~(\ref{T2corrected}). We find that for $R_{se}\gg R_{sd}$ the maximum value
of $dP_i/d\omega/\sqrt{T_2}$ is given by $dP_i/d\omega/\sqrt{T_2}|_{\max}=k R_{se}^{-1/2}$, where $k=1.3$. This result is also verified with the full numerical density matrix model. With $R_{se}= n \bar{v}\sigma_{se}$, the optimal sensitivity of a scalar alkali-metal magnetometer is given by
\begin{equation}
\delta B_{\min }=\frac{0.77}{\gamma} \sqrt{\frac{\bar{v}\sigma_{se}(1+\eta^{-1/2})}{V}}.
\end{equation}

Hence we find that for a scalar magnetometer the fundamental sensitivity is limited by the rate of spin-exchange collisions even though the resonance linewidth can be much smaller than the spin exchange rate. Numerically for $\sigma_{se}=1.8\times 10^{-14}$ cm$^{2}$ and $\eta=0.8$ we find that $\delta B_{\min}=0.9$ fT/Hz$^{1/2}$ for an active volume of 1~cm$^3$. Using back-action evasion techniques it is possible to make the light shift and photon shot noise contributions negligible, but this only improves the sensitivity to 0.6 fT/Hz$^{1/2}$ for 1 cm$^3$ volume.

If total noise is limited by photon shot noise or by technical sources of rotation noise, as was the case in our experiment (see Fig. \ref{fig:results}), the sensitivity is optimized by maximizing the slope on resonance
$dP_i/d\omega$. Using the same optimization procedure using modified BE and varying $R_{op}$ and $B_1$ one can obtain $dP_i/d\omega|_{\max}=1.2 R_{se}^{-3/4} R_{sd}^{-1/4}$. In this case the maximum slope is increased from $R_{se}^{-1}$ scaling that one would obtain with a spin-exchange-broadened resonance from Eq. (\ref{max_slope}). Therefore, light narrowing is useful in reducing the noise in scalar magnetometers limited by photon shot noise or $1/f$ noise, with a maximum sensitivity gain on the order of $(R_{se}/R_{sd})^{1/4}$, which is equal to about 10 for K atoms.

One can also use Eq.~(\ref{unoptnoise}) to estimate the best sensitivity possible under our actual experimental conditions. In this case the number of absorption length on resonance $N_{ab}$ is not optimal and the spin relaxation of K atoms has additional contribution from collisions with buffer gas and diffusion to the walls. For our parameters corresponding to Fig.~\ref{fig:results} ($N_{ab}=2.5$, $R_{se}=8700$~s$^{-1}$, $R_{pr}\sim 100$~s$^{-1}$, $R_{sd}+R_{pr}\sim 130
~$s$^{-1}$,  $V \sim 1.5$~cm$^3$ and $\eta=0.24$), including losses
in collection of probe light after the cell), we get optimal sensitivity from Eq.~(\ref{unoptnoise}) of 7 fT/Hz$^{1/2}$, dominated by photon shot noise. This compares well with the measured photon shot noise level corresponding to 7 fT/Hz$^{1/2}$ in each channel.
The sensitivity could be improved by increasing the resonance optical depth of the vapor.
% dP/dw=0.0003, R = 2000-3000
% matches linewidth and RF strength
%about 10% less than maximum slope

\section{Conclusion}
In this paper we have systematically analyzed the sensitivity of a scalar
alkali-metal magnetometer operating in the regime where the relaxation is dominated by spin-exchange collisions.
We demonstrated experimentally magnetic
field sensitivity below 10 fT~Hz$^{-1/2}$ with an active
volume of 1.5 cm$^3$, significantly improving on previous sensitivities obtained for  scalar atomic magnetometers
and opening the possibility for further miniaturization of such sensors.

We considered the effects of rf broadening in the presence of SE relaxation and developed a simple analytic model
based on Bloch equations with a $T_2$ time that depends on rf excitation. The results of the model have been validated
against a complete numerical density matrix calculation and experimental measurements.
We showed that the fundamental sensitivity limit for a scalar alkali-metal
magnetometer with a 1 cm$^3$ measurement volume is on the order of
0.6-0.9~fT~Hz$^{-1/2}$. In this case a reduction of resonance linewidth by optical pumping of atoms into a stretched
state does not lead to an improvement of fundamental sensitivity limit.

It is interesting to compare the scaling of the optimal magnetic
field sensitivities in various regimes.  It was shown in
\cite{KominisRomalisNature} that near zero field in the SERF
regime the sensitivity scales as $\sigma_{sd}^{1/2}$, while for an
rf  magnetometer operating in a finite field it scales as
$(\sigma_{se} \sigma_{sd})^{1/4}$ \cite{rfmagn}. In contrast,
here we find that the fundamental sensitivity limited by spin projection noise for a scalar magnetometer
in a finite field scales as $\sigma_{se}^{1/2}$, i.e. there is no
significant reduction of SE broadening for optimal conditions. Since
$\sigma_{se}$ is similar for all alkali metals, one can expect a
similar sensitivity for a Cs or Rb magnetometer. On the other hand,
if one is limited by the photon shot noise or technical sources of
optical rotation noise, which is often the case in practical systems, the
magnetometer sensitivity is determined by the slope of the dispersion curve. In this case
it is improved in the light-narrowing regime because the slope of the
dispersion resonance scales as $\sigma_{se}^{-3/4}
\sigma_{sd}^{-1/4}$, instead of $\sigma_{se}^{-1}$  for the case of spin-exchanged broadened resonance.  We expect similar relationships, with different numerical factors, to hold for
atomic clocks operating on the end transitions, since $T_2$ in that
case is given by an equation similar to Eq.~(\ref{T2corrected})
\cite{Jau04}. The analytical approach developed in this paper can be easily adapted to other
alkali atoms by modifying the coefficients in Eq.~(\ref{T2corrected}). This work was
supported by an ONR MURI grant.
\bibliography{scalarbib}
\end{document}